\documentclass[letterpaper]{article} 
\usepackage{aaai23}  
\usepackage{times}  
\usepackage{helvet}  
\usepackage{courier}  
\usepackage[hyphens]{url}  
\usepackage{graphicx} 
\usepackage{natbib}  
\usepackage{caption} 
\usepackage{algorithm}
\usepackage{algorithmic}

\usepackage{newfloat}
\usepackage{listings}
\DeclareCaptionStyle{ruled}{labelfont=normalfont,labelsep=colon,strut=off} 
\floatstyle{ruled}
\newfloat{listing}{tb}{lst}{}
\floatname{listing}{Listing}
\title{Identifying and Characterizing Behavioral Classes of Radicalization within the QAnon Conspiracy on Twitter}
\author {
    Emily L. Wang,\textsuperscript{\rm 1,2}
    Luca Luceri,\textsuperscript{\rm 1}
    Francesco Pierri,\textsuperscript{\rm 1,\rm 3}
    Emilio Ferrara\textsuperscript{\rm 1}\\
}
\affiliations {
    \textsuperscript{\rm 1} Information Sciences Institute, Viterbi School of Engineering, University of Southern California, CA, USA \\
    \textsuperscript{\rm 2} Northwestern University, IL, USA \\
    \textsuperscript{\rm 3} Politecnico di Milano, Milan, Italy \\
    emilywang2024@u.northwestern.edu, lluceri@isi.edu, fpierri@isi.edu, emiliofe@usc.edu
}

\begin{document}

\maketitle

\begin{abstract}
Social media provide a fertile ground where conspiracy theories and radical ideas can flourish, reach broad audiences, and sometimes lead to hate or violence beyond the online world itself. 
QAnon represents a notable example of a political conspiracy that started out on social media but turned mainstream, in part due to public endorsement by influential political figures.
Nowadays, QAnon conspiracies often appear in the news, are part of political rhetoric, and are espoused by significant swaths of people in the United States. 
It is therefore crucial to understand how such a conspiracy took root online, and what led so many social media users to adopt its ideas.
In this work, we propose a framework that exploits both social interaction and content signals to uncover evidence of user radicalization or support for QAnon.
Leveraging a large dataset of 240M tweets collected in the run-up to the 2020 US Presidential election, we define and validate a multivariate metric of radicalization.
We use that to separate users in distinct, naturally-emerging, classes of behaviors associated with radicalization processes, from self-declared QAnon supporters to hyper-active conspiracy promoters. 
We also analyze the impact of Twitter's moderation policies on the interactions among different classes: we discover aspects of moderation that succeed, yielding a substantial reduction in the endorsement received by hyperactive QAnon accounts. But we also uncover where moderation fails, showing how QAnon content amplifiers are not deterred or affected by the Twitter intervention.
Our findings refine our understanding of online radicalization processes, reveal effective and ineffective aspects of moderation, and call for the need to further investigate the role social media play in the spread of conspiracies.
\end{abstract}

\section{Introduction}

Social media platforms are becoming increasingly instrumental in the spread of conspiracy theories that originate from and are reinforced by close-knit, niche online communities \cite{papakyriakopoulos2020spread, stecula2021social}. In these environments, extreme opinions and fringe ideas can gather momentum and eventually gain traction on mainstream media \cite{tollefson2021tracking}. While some conspiracy theories pose relatively little harm, others create distrust towards institutions or the government, which can in turn result in harm or violence offline.

One prominent example in recent times is QAnon, a conspiracy that supports a range of extreme and uncorroborated ideas---suggesting that a cabal of Satan-worshiping, cannibalistic pedophiles control politics and media in the US, with former President Trump waging war against it.\footnote{https://www.nytimes.com/article/what-is-qanon.html} Under the umbrella of QAnon's theories we find the ``Pizzagate''
conspiracy,\footnote{https://www.bbc.com/news/blogs-trending-38156985} a pillar of QAnon since the 2016 US Presidential Election that claims Hillary Clinton was running a child sex ring; the ``Obamagate'' conspiracy,\footnote{https://www.voanews.com/a/usa\_us-politics\_what-obamagate/6189342.html} which accuses former President Obama of spying on former President Trump; and many more recent conspiracies surrounding the COVID-19 pandemic \cite{ferrara2020what,pierri2022one,nogara2022disinformation} and the 2020 US Presidential election \cite{abilov2021voterfraud2020, suresh2023tracking}. Due to its connection with major political actors, QAnon broadly influences the political sphere by spreading harmful propaganda from fringe online groups to mass social media \cite{hannah2021qanon}. In the 2020 US Congressional elections, 106 candidates endorsed QAnon ideas,\footnote{https://www.mediamatters.org/qanon-conspiracy-theory/here-are-qanon-supporters-running-congress-2020} and 2 of them obtained House of Representatives seats, whereas in the upcoming 2022 US Congressional elections, 65 candidates have expressed support for QAnon.\footnote{https://www.mediamatters.org/qanon-conspiracy-theory/here-are-qanon-supporters-running-congress-2022} 

Beyond political influence, QAnon support at times also escalates and turns into real-world acts of violence \cite{luceri2021social}. 
One of the most notorious is the January 6th, 2021 attack on the US Capitol, where at least 34 QAnon followers, including the self-proclaimed ``QAnon Shaman,'' participated.\footnote{https://www.voanews.com/a/usa\_capitol-riot-exposed-qanons-violent-potential/6203967.html} 
In fact, QAnon’s tendency for violence had already alarmed experts before this attack. In July 2020, the \textit{Combating Terrorism Center at West Point} described QAnon as ``\textit{a public security threat with the potential in the future to become a more impactful domestic terror threat}'' and detailed notable criminal cases related to the conspiracy, which include charges of assault, terrorist threats, and murder \cite{amarasingam2020qanon}. More recently, the FBI stated in a threat assessment\footnote{https://www.cnn.com/2021/06/14/politics/fbi-qanon-warning-to-lawmakers/index.html} in June 2021 that domestic violent extremists who are also self-identified QAnon adherents believe ``\textit{they have an obligation to change from serving as `digital soldiers’ towards engaging in real-world violence.}'' Existing literature confirms the association between conspiracy belief and endorsement of political violence \cite{vegetti2022belief, jolley2020pylons}. In fact, according to \citet{imhoff2021resolving}, adopting beliefs in a hypothetical conspiracy theory similar to QAnon resulted in an increased willingness to commit illegal, non-normative political acts like violence. 

With these concerns in mind, it is increasingly important to understand how individuals may become radicalized online, starting from social media platforms---where users are first exposed to and then engage with QAnon ideas---and later shifting to real-world action. Recent research characterized top spreaders in a QAnon-related group on the niche platform Voat \cite{papasavva2021qoincidence}; other studies analyzed radicalization pathways \cite{fabbri2022rewiring} to conspiracy on Youtube \cite{ribeiro2020auditing} and Reddit \cite{phadke2022pathways}. 
Existing work on QAnon-related activity on Twitter explored the temporal dynamics of engagements with, and diffusion of, QAnon content at the aggregate level \cite{cunningham2022network, xu2022network, sharma2022characterizing}. These contributions employ binary indicators (e.g., checking whether a user shared a QAnon keyword at least once) to identify users endorsing QAnon narratives. 
However, our understanding of user-based adoption of a conspiracy such as QAnon via social media is still lacking.
Therefore, in this work, we aim at identifying and characterizing a diverse set of behaviors tied to users' engagement with QAnon theories. To this aim, we propose a set of continuous metrics to capture signals of radicalization on Twitter and uncover distinctive behaviors of radicalized users.

\subsection{Research Questions}
To unveil signals tied to users’ radicalization within the QAnon conspiracy and discern behavioral differences among users, we focus on the following research questions:
\begin{itemize}
    \item[\textbf{RQ1}] Do we observe distinct classes of behaviors across different radicalization metrics? 
    \item[\textbf{RQ2}] If so: How do behavioral classes of radicalization differ and interact with one another?
\end{itemize}


To address these questions, we leverage a dataset of over 240 million election-related tweets collected between June and September 2020 in the run-up to the US Presidential election. We measure signals of radicalization in users’ tweets and profiles as well as through their social connections. These signals form a basis to identify and characterize radicalization processes and behaviors.

\subsection{Contributions of this work}
The framework outlined above led us to the following fundamental contributions:
\begin{itemize}
    \item We present a framework for measuring users’ signals of radicalization within the QAnon conspiracy, both in explicit content production as well as through social interactions. 
    We provide a suite of four continuous metrics of radicalization, which allow us to separate different facets of engagement with QAnon. These metrics are agnostic to the platform and group under analysis, thus the framework can be generalized to other scenarios and platforms. 
    
    \item We suggest that radicalization processes should be modeled across multiple dimensions: one single dimension cannot capture the complex facets of radicalization.
    Leveraging clustering techniques, we discover six distinct behavioral classes of radicalization, each associated with different behaviors. 
    The main archetypes of radicalized users include conspiracy amplifiers, self-declared supporters, and hyper-active promoters. 
    The three most radicalized classes comprise of a significant proportion ($\sim$9.4\%) of the users in the dataset under analysis. 
    
    \item We observe that users in the most radicalized classes tend to share less reliable URLs and are significantly more likely to be suspended by Twitter. Hyperactive promoters are most persistent in sharing QAnon content over time and engage with a large variety of QAnon topics. 
    Amplifiers and hyper-active promoters tend to rebroadcast content originating from their own groups  significantly more than expected. Comparing the interactions between QAnon accounts and other users, before and after Twitter intervention against QAnon, we find that engagements with hyper-active QAnon promoters are reduced substantially, whereas QAnon amplifiers were not significantly affected by Twitter moderation.
\end{itemize}

\section{Related Work}

\subsection{QAnon on Social Media}

Although the QAnon conspiracy had its genesis on niche platforms like 4chan and 8kun, its advocates quickly migrated to mainstream media like Youtube and Reddit, and appeared in online news channels after participating in a Trump rally \cite{de2020tracing}. 
Since then, QAnon has increasingly percolated into the mainstream. Compared to other far-right groups with fringe ideas, QAnon adherents share and consume significantly more mainstream conservative content \cite{zihiri2022qanon}. \citet{xu2022network} find that QAnon has been evolving into a wider umbrella of beliefs across the news, US politics, and COVID-19, gaining more relevance as a result. In fact, in the run-up to the 2020 US Presidential election, the majority of active US Twitter users, both from the political left and right, had interacted with QAnon accounts \cite{sharma2022characterizing}. 

In response to real-world violence associated with QAnon, Twitter, Youtube, and Facebook introduced intervention measures aimed at demoting accounts and pages associated with the conspiracy.\footnote{https://www.pbs.org/newshour/nation/youtube-follows-twitter-and-facebook-with-qanon-crackdown} These measures affected the use of QAnon-related keywords and hashtags. In fact, \cite{cunningham2022network} contrasted the diffusion of QAnon content before and after the Twitter intervention, finding a substantial decline in tweets using the hashtag \#QAnon after Twitter action, although similar content was pushed by new central actors. 
Another network-based approach 
concluded that bots played a mixed role in propagating and countering the spread of QAnon conspiracy theories, and identified a large group of users who do not have a clear stance on QAnon as a possible target for mitigation efforts \cite{xu2022network}. 
In contrast to these contributions, we study the engagement with QAnon content at the user level to examine multiple dimensions of radicalization. 


\subsection{Radicalization Theories} 

Many theories in the literature attempt to explain the process of radicalization, which typically includes motivational, ideological, and social factors. For example, the ``3N model" \cite{kruglanski2014psychology} refers to three important components of  radicalization: 1) \emph{Need}: a goal to which an individual is highly committed, 2) \emph{Narrative}: justification of violence as an appropriate mean to pursue goals, and 3) \emph{Network}: a community that echoes extreme ideology while isolating individuals from opposing ideas \cite{kruglanski2014psychology}. This model was validated empirically in four cultural settings \cite{belanger2019radicalization} and recently interpreted in the context of conspiracy theories \cite{kruglanski2022terrorism}.

Out of the 3Ns, \emph{Network} is most suitable to study with social media data. Other radicalization models also include a social component like \emph{Network}. For instance, in Sageman’s Four Prongs, the notion of “mobilization through networks” involves confirming one's ideas and interpretation of events with other radicalized people \cite{sageman2008strategy}. Also, the RECRO model of internet-mediated radicalization \cite{neo2019internet} includes a \emph{Connection} phase, where individuals build mutual trust with members of the community while distrusting opposing voices. The widely-recognized influence of communities on radicalization is no surprise: sympathizers attracted to radical ideas isolate themselves and engage only with those who support their views, which causes their ideas to become more extreme \cite{yardi2010dynamic}. With the aim of reducing the stigma around extremist ideas, some believers
self-disclose their allegiance on social media, fostering a strong group identity
\cite{torok2013developing}. These theoretical works are particularly relevant for this paper, as they inform our definition of community-based signals of radicalization and help explain some of our results like users' interactions, self-disclosure, and engagement with QAnon.

\subsection{Signals and Stages of Radicalization}

Many approaches have been proposed to quantify and detect radicalization. \citet{rowe2016mining} defined signals of radicalization as using extremist language (i.e., specific keywords) and retweeting incitement messages generated by
known extremist accounts. 
Focusing on the roots of radicalization, \citet{fernandez2018understanding} quantified micro- (individual), meso- (group/community), and macro- (global) roots by looking at the use of radicalized terminology in the original post, re-shares, and links published in users' timelines,
respectively. 
In addition to the analysis of shared keywords, \citet{nouh2019understanding} extracted psychological traits from text, such as tendencies to use certain personal pronouns, as well as behavioral characteristics, such as frequency of tweets and followers/following ratio. 

Although research surrounding Internet-mediated radicalization grew largely in response to violent extremists, such as Al Qaeda and ISIS, promoting propaganda on social media \cite{neo2019internet}, recent work has started to investigate radicalization in
conspiracy theories and far-right movements. \citet{phadke2022pathways} quantified the first three stages of the RECRO model (i.e., reflection, exploration, and connection) for $\sim$36,000 users on Reddit and found that they correlated with temporal patterns of engagement with conspiracy discussions.
In \cite{schulze2022far}, authors manually analyzed a sample of Telegram posts within three far-right movements active in Germany and extracted different indicators of radicalization, including conspiracy narratives, anti-elitism, calls for political participation and activism, and support for violence. Differently from these studies, we focus on a unique conspiracy group to define and validate a multivariate metric of radicalization. 
While the proposed framework can be extended to different social media platforms and fringe communities, we start with QAnon to generate actionable and targeted insights.

\section{Methodology}
 
In this section, we describe the data employed in the analysis and we detail the methodology to extract the designed signals of radicalization. The first two signals are content-based: \textit{(i)} QAnon content in shared tweets and \textit{(ii)} QAnon content present in users' profiles; the other two signals are community-based: \textit{(iii)} retweets to, and \textit{(iv)} lexical similarity with a seed of QAnon promoters. To capture these metrics,\footnote{Note that throughout the paper, \emph{signals} and \emph{metrics} will be used interchangeably} we leverage a set of QAnon-related keywords and Web domains, alongside a set of seed users persistently engaged with the diffusion of QAnon content, henceforth referred to simply as \emph{persistent QAnon users}. 

\subsection{Data Collection}
We employ a dataset of election-related tweets collected using Twitter’s streaming API service in the run-up to the 2020 US election \cite{chen2022election2020}. We focus on the same time period analyzed by \citet{sharma2022characterizing}, namely from June 20, 2020 to September 6, 2020, as we aim to compare our methodology and findings. In this time period, we observe 242,087,331 tweets (including original tweets, replies, retweets, and quote retweets) shared by 10,392,492 unique users. 
The majority of these users only posted a handful of tweets during the period of observation, from which limited insight can be drawn.
Hence, we consider only tweets by 1,207,646 most active users who shared at least 20 tweets in the collected dataset. This threshold ensures sufficient activity over the time period for each user and reduces the size of the retweet network required for inferring political leaning as in \cite{sharma2022characterizing}. We further split the data into eleven weeks to study users’ activity over time.


\subsection{Measuring Content-Based Signals}

We identify tweets sharing QAnon content using a list of keywords frequently associated with the QAnon conspiracy and previously published by \citet{sharma2022characterizing}. Furthermore, we identify QAnon-associated URLs using a list of 324 Web domains, which were manually flagged for sharing QAnon content during our observation period \cite{hanley2022no}. Note that, as of July 2022, some of these Websites are no longer active or have shifted to promoting general conspiracy content. A sample of keywords and domains are shown in Table \ref{table1}.

\begin{table}
\centering
\begin{tabular}{c|c}
\hline
\textbf{Keyword}  & \textbf{Domain}              \\ \hline
wwg1wga           & qanon.pub                    \\ \hline
\#obamagate       & qdrop.pub                    \\ \hline
\#qanon           & operationq.pub               \\ \hline
\#savethechildren & x22report.com                \\ \hline
deepstate         & wwg1wga.martingeddes.com     \\ \hline
thegreatawakening & theqpatriothub.weebly.com    \\ \hline
wgaworldwide      & voat.co                      \\ \hline
\#qarmy           & wg1wga.com                   \\ \hline
\#pizzagate       & qcon.live                    \\ \hline
\#taketheoath     & thegreatawak-eningsummit.com \\ \hline
\end{tabular}
\caption{Example of QAnon keywords and domains}
\label{table1}
\end{table}

To measure the prevalence of QAnon-related content shared by users, we examine the presence of QAnon keywords and Web domains in every tweet of our dataset. The usage of this set of keywords and domains naturally embodies an engagement with QAnon theories and minimizes the likelihood of false positives (i.e., misclassified messages). Rather than using a binary variable, to define a user as radicalized or not, 
as well as to study their engagement with the conspiracy over time,
we provide a continuous scale capturing the degree of users' involvement in QAnon.

To quantify to what extent users share QAnon content in their tweets,
we define the metric $QC_{tweets}$ as follows:

\[
QC_{tweets} = \frac{\textrm{No. of QAnon keywords + no. of QAnon URLs }}{\textrm{No. of tweets}}.  
\]


$QC_{tweets}$ is greater than 1 whenever users include more instances of QAnon-related content (either keywords or URLs) in their tweets than the total number of their tweets. Moreover, we refer to \emph{self-drafted $QC_{tweets}$}, when considering only original content generated by the users, i.e., original tweets, replies, and quotes. 

Similarly, we measure the prevalence of QAnon-related content in a user’s profile description, and define the metric $QC_{profile}$ as follows: 

\[
QC_{profile}  = \frac{\textrm{No. chars in QAnon keywords/URLs in profile}}{\textrm{No. chars in profile}}.  
\]

As users might change their profile description over time, we consider the most frequent text appearing in the dataset during the period of analysis. Empty (zero-character) profiles are assigned a default value of 0. 

\subsection{Identifying Persistent QAnon Users}

In addition to measuring signals of radicalization from QAnon-related content, we aim to understand and quantify how users interact with the QAnon community as a whole. We thus define a seed group of \emph{persistent QAnon users} by strictly filtering for users who satisfy both conditions below:  
\begin{itemize}
    \item \emph{self-drafted} $QC_{tweets} > 0$ over the entire time period -- to ensure that these users produced original QAnon content.
    \item $QC_{tweets} > 0$ and more than 10 tweets for each of the first five weeks of the period of observation (June 20 - July 24, which encompasses the time period before substantive Twitter action against QAnon) -- to avoid selecting users who do not show consistent QAnon-related activity.
\end{itemize}

When calculating $QC_{tweets}$ over a single week, we observe that some users shared a limited number of tweets. Thus, we compute $QC_{tweets}$ only for users with more than 10 tweets shared every week. This ensures a fairer comparison between users, as users with very few tweets shared are more likely to have inflated $QC_{tweets}$---a user with one QAnon keyword out of one tweet is substantially different from a user with ten QAnon keywords out of ten tweets, despite the ratio being the same. 

\begin{figure}[t]
\centering
\includegraphics[width=.9\linewidth]{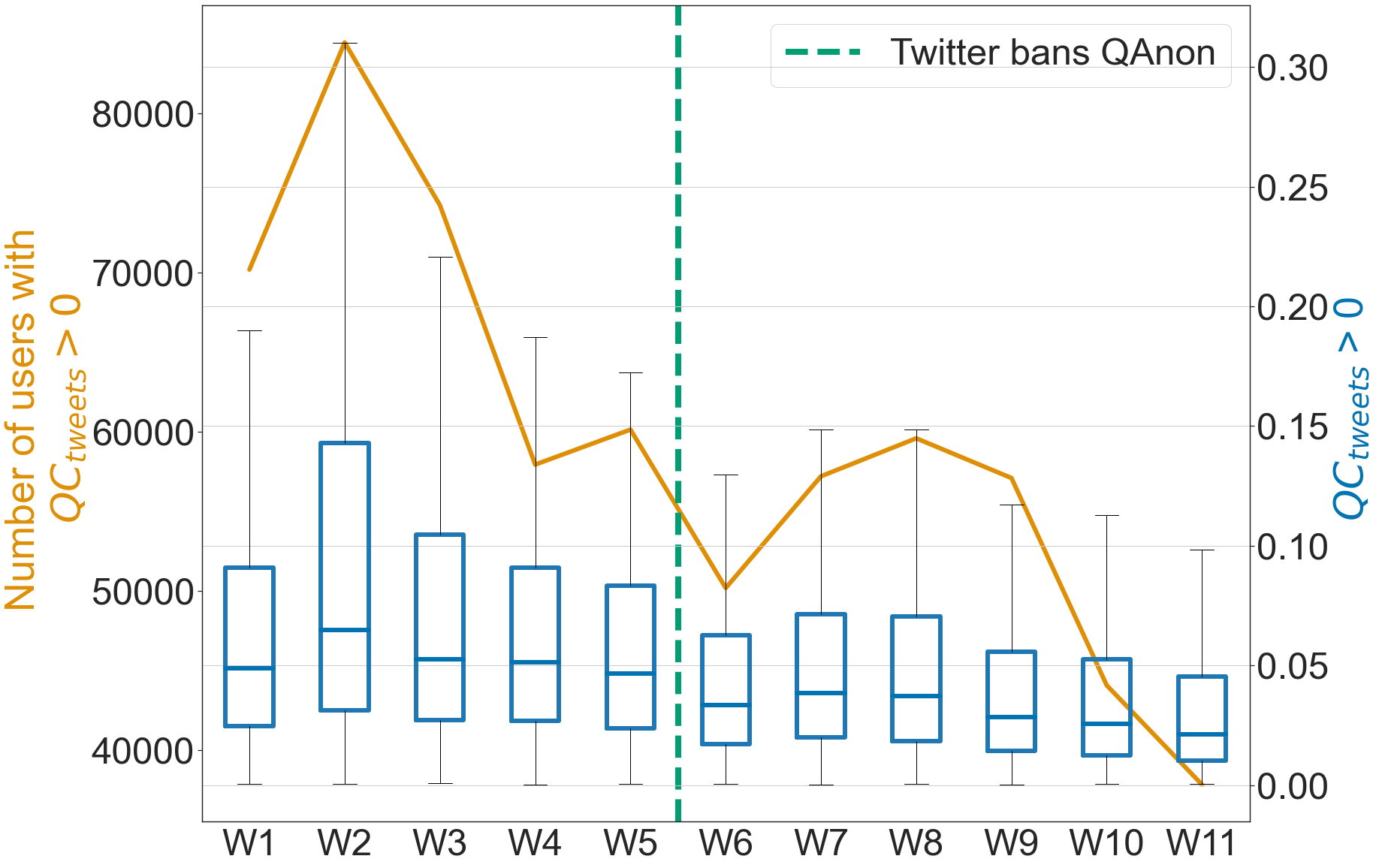}
\caption{
Number of users with $QC_{tweets} > 0$ (left y-axis) and the distribution of their $QC_{tweets}$ (right y-axis) for each week of the dataset 
}
\label{figure1m}
\end{figure}

\begin{table*}[t]
\begin{tabular}{|l|l|l|l|l|}
\hline
& \textbf{Condition}                                                                                                                                                                         & \begin{tabular}[c]{@{}l@{}}\textbf{Total}\\ \textbf{users}\end{tabular} & \begin{tabular}[c]{@{}l@{}}\textbf{Left-leaning}\\ \textbf{users}\end{tabular} & \begin{tabular}[c]{@{}l@{}}\textbf{Proportion} \\ \textbf{left-leaning}\end{tabular} \\ \hline
\begin{tabular}[c]{@{}l@{}}Previous work \\ (Sharma 2022)\end{tabular} & \begin{tabular}[c]{@{}l@{}}Shared at least one QAnon keyword \\ in self-drafted tweet\end{tabular}                                                                                & 92,065                                                & 7,661                                                        & 8.32\%                                                             \\ \hline
Baseline approach                                                      & $QC_{tweets} > 0$ over the entire time period                                                                                                                             & 227,721                                               & 25,794                                                       & 11.3\%                                                             \\ \hline
Proposed approach                                                      & \begin{tabular}[c]{@{}l@{}}\emph{Self-drafted} $QC_{tweets} > 0$ over the entire time period \\ and $QC_{tweets} > 0$ in each week preceding Twitter action\end{tabular} & 6,536                                                 & 50                                                           & 0.765\%                                                            \\ \hline
\end{tabular}
\caption{Comparison of different conditions for defining \emph{persistent QAnon users}}
\label{table2}
\end{table*}

Figure \ref{figure1m} depicts the total number of users engaged with QAnon content and the distribution of their $QC_{tweets}$ in the observation period. Both measures exhibit a significant drop in the weeks following the intervention by Twitter, which starting from July 21, 2020, enacted restrictions on the spread of QAnon content, limiting its appearance in trends and search, and removing accounts or content related to QAnon from recommendations.\footnote{https://www.nbcnews.com/tech/tech-news/twitter-bans-7-000-qanon-accounts-limits-150-000-others-n1234541} \footnote{Although substantive action was announced on July 21, 2020, certain QAnon influencers reported limits in early July.} Previous work indicated that users evaded these regulations by decreasing their usage of certain QAnon hashtags and keywords while maintaining a high volume of tweets and pivoting to other keywords \cite{sharma2022characterizing}.  Therefore, we only consider the five weeks before the ban for defining our group of \emph{persistent QAnon users} in order to also capture the users who attempted to evade Twitter action.


\subsection{Validating Persistent QAnon Users}

In previous work \cite{sharma2022characterizing}, QAnon users were identified by using two criteria: (1) they shared at least one tweet containing QAnon keywords and (2) they were inferred as right-leaning. 
The political leaning of users was estimated by considering links to media outlets in tweets and leveraging the classification of web domains provided by Allsides.\footnote{\url{http://allsides.com}} Label propagation on the retweet network was then used to infer the political ideology of the remaining users. A manual check by one author on a random sample of both left- and right-leaning users showed that the former group contained mostly false positives (identified by checking if the user's QAnon-related tweets were all ridiculing QAnon), whereas the latter were indeed QAnon supporters. Thus, we infer political leanings using the same approach as in \cite{sharma2022characterizing}, and we consider the proportion of left-leaning users as a proxy for users who only reference QAnon by chance. 

Table \ref{table2} reports the proportion of inferred left-leaning users who satisfy a different set of conditions for defining the set of \emph{persistent QAnon users}. It can be noticed that imposing stricter conditions, on both temporal and content engagement, reduces the proportion of left-leaning users by more than an order of magnitude. This suggests that our selected \emph{persistent QAnon users} are indeed advocates of QAnon rather than just mentioning it by chance. As a robustness check, we manually examined the Twitter timelines of the 50 left-leaning persistent users and found that 10 were incorrectly inferred as left-leaning and QAnon promoters, while the other 40 were highly dedicated to attacking or trying to convert QAnon users. We also examined the Twitter timelines of a random sample of 50 inferred right-leaning persistent users and found that every user was indeed a QAnon promoter. Thus, consistently with \cite{sharma2022characterizing},
we consider only the 6,486 inferred right-leaning users as \emph{persistent QAnon users}. 
 
\subsection{Measuring Community-Based Signals}

The ``connection'' or ``network'' phase of radicalization theories describes how individuals' worldview changes as they interact with and are influenced by a like-minded community \cite{sageman2008strategy, kruglanski2014psychology, torok2013developing, neo2019internet}. As individuals interact with others who share similar ideology \cite{mcpherson2001birds}, they may start conforming to the social group and mimic the specific vocabulary used by their peers. We devise two community-based metrics that might uncover signals of radicalization. Using the group of \emph{persistent QAnon users} as a reference, we measure to what extent users engage with and communicate similarly to QAnon accounts.

Considering retweets as a form of endorsement, in line with recent work \cite{metaxas2015retweets, stella2018bots}, we measure the proportion of users’ re-sharing of the content generated by \emph{persistent QAnon users}, and define $C_{retweets}$ as follows:

\[
C_{retweets} = \frac{\textrm{No. of retweets to persistent QAnon users}}{\textrm{No. of retweets}}.  
\]


To measure to what extent users are similar to the \emph{persistent QAnon users}, we also define  a lexicon based on the content they shared. Specifically, to quantify users' lexical similarity with \emph{persistent QAnon users}, we tokenize every tweet in the dataset 
with the NLTK Python Library \cite{bird2009natural}, which converts tweets into tokens (i.e. words, punctuation, emojis, and hashtags) and removes handles. We compare token frequencies between the self-drafted tweets generated by \emph{persistent QAnon users} (combined into a single corpus) and the self-drafted tweets shared by every other user in the dataset (combined into another corpus) using a weighted log-odds ratio with uniform Dirichlet priors of 0.01 \cite{monroe2008fightin} and we select the top 0.5\% tokens with the highest weighted log-odds ratios. Higher weighted log-odds ratios indicate tokens frequent in \emph{persistent QAnon users}’ tweets compared to the background 
(other users' messages). We chose a cutoff of 0.5\% to use a lexicon as large as possible without compromising the separation between the distributions of lexical overlap between \emph{persistent QAnon users} and other users. 

\begin{figure}[t]
\centering
\includegraphics[width=.9\linewidth]{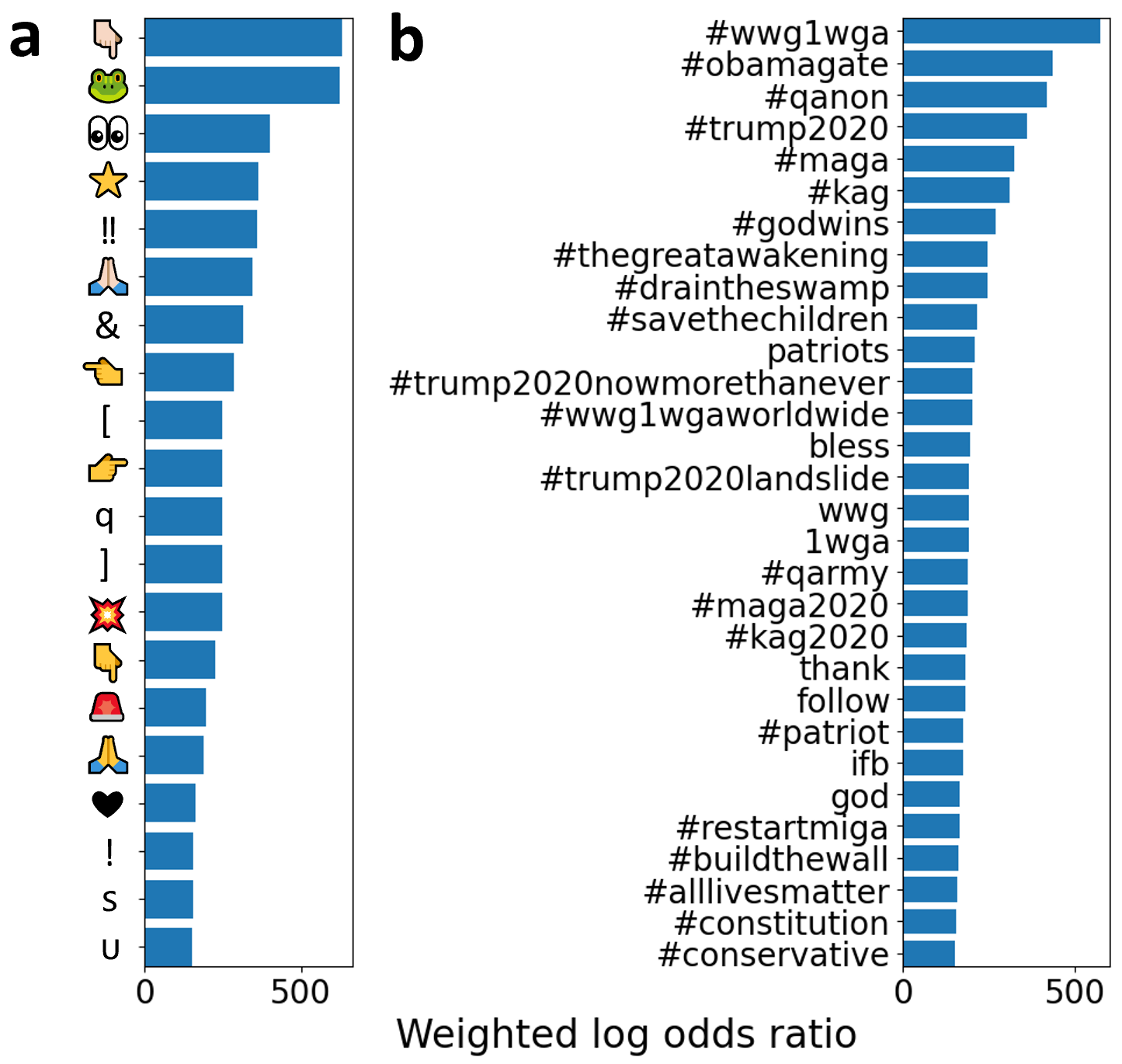}
\caption{
(a) Top 20 one- or two-character tokens of \emph{persistent QAnon users} 
(b) Top 20 tokens of \emph{persistent QAnon users} after filtering out stopwords, punctuation, and words with less than 3 characters or no alphabetic characters
}
\label{figure3m}
\end{figure}

The top 20 one- or two-character tokens are shown in Figure \ref{figure3m}\textbf{a}. We observe several pointing and praying emojis as well as the \emph{frog} emoji, which members of the alt-right began adding to their Twitter handles in solidarity with white nationalism \cite{zannettou2018origins, hine2017kek}. We take the list of 17,779 tokens and filter out stopwords, punctuation, and words with less than 3 characters or no alphabetic characters like emojis to yield a \emph{persistent QAnon users}’ lexicon of 15,632 tokens, which includes 7,474 hashtags. The top 20 tokens after this filtering are displayed in Figure \ref{figure3m}\textbf{b}. 
Common words and hashtags are related to God, Trump, QAnon, and \emph{following} activity (“follow”, “ifb” = “I follow back”), which may indicate the presence of ``follow-back" communities (see Section \emph{Behavioral Classes of Radicalization (RQ2)} for more details). As this lexicon is much larger than the list of 30 keywords (described before), it can capture more subtle linguistic cues of QAnon accounts, including style and lingo. Moreover, as it is highly challenging to exhaustively list all QAnon-related keywords, given that some are less popular or secluded to particular circles, this method helps us find lesser-known keywords---for example, ``\#restartmiga" refers to ``Restart," an Iranian subgroup of QAnon.\footnote{https://foreignpolicy.com/2020/07/15/qanon-goes-to-iran/}



We then tokenize each individual user’s self-drafted tweets to measure lexical similarity $C_{lexical}$ with \emph{persistent QAnon users}’ lexicon,
quantified as follows: 

\[
C_{lexical} = \frac{\textrm{No. of persistent QAnon users' lexicon words}}{\textrm{No. of words}}.  
\]

 
\section{Results}


\subsection{Signals of Radicalization (RQ1)}

\begin{figure}[t]
\centering
\includegraphics[width=.9\linewidth]{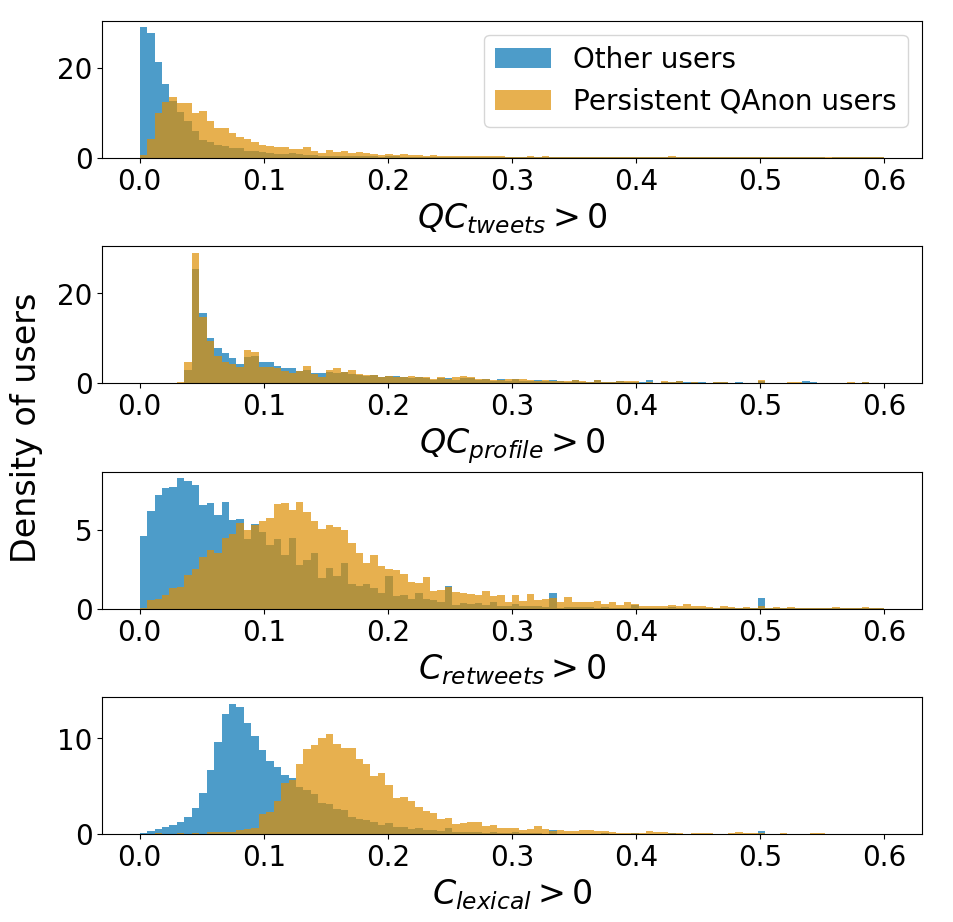}
\caption{
Distribution of the four metrics ($QC_{tweets}$, $QC_{profile}$, $C_{retweets}$, and $C_{lexical}$) for \emph{persistent QAnon users} and other users. Only values $> 0$ are considered. 
}
\label{figure-1r}
\end{figure}

\begin{figure}[t]
\centering
\includegraphics[width=.75\linewidth]{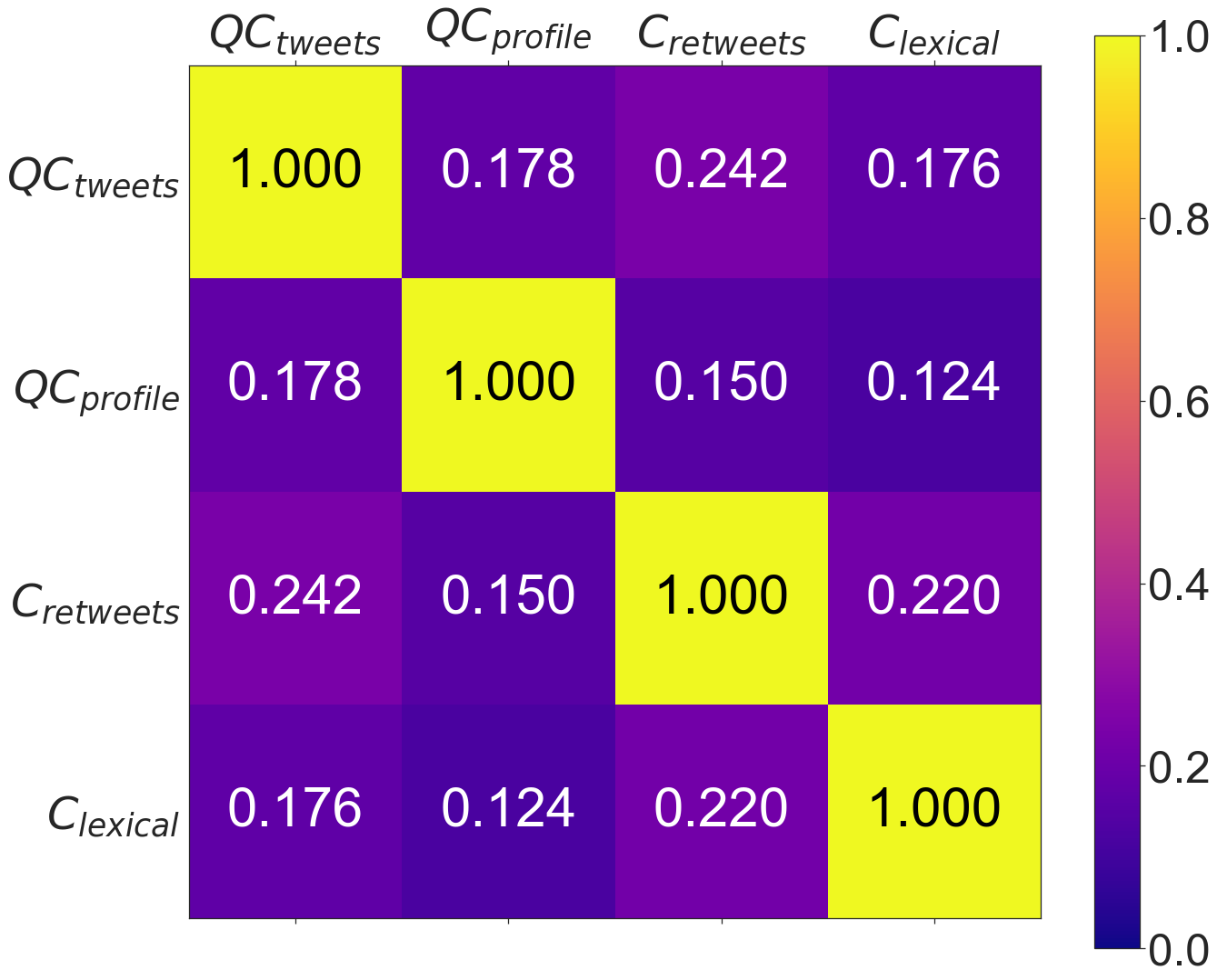}
\caption{
Pairwise Pearson correlation between the four signals of radicalization
}
\label{figure0r}
\end{figure}

To address RQ1, we compute the four signals of radicalization and we examine their distribution over the entire population under observation. Figure \ref{figure-1r} depicts the distribution of the four metrics, separating \emph{persistent QAnon users} from other users in our dataset. 
We focus on the latter group to study the radicalization process, and set aside the \emph{persistent QAnon users} group for future studies since this group is already committed to the conspiracy.
From Figure \ref{figure-1r}, it can be noticed that most of the distributions of the four signals of radicalization are right skewed, with most of their probability mass in the range [0,0.2]. 
The distribution of all metrics except $QC_{profile}$ for \emph{persistent QAnon users} and other users are statistically different (p-value $< 0.001$).
We observe that there is not a unique signal that distinctively characterizes \emph{persistent QAnon users},
and that can be used as a term of comparison to measure
the degree of radicalization of other users.

To further investigate whether multiple signals are needed to model radicalization processes, we examine the correlation among the four metrics. The idea is that if the metrics are correlated, we might use one metric (or their combination) as a measure of radicalization.
Contrarily, if they are not correlated, we can assume that each metric conveys an independent signal of radicalization. In Figure \ref{figure0r}, where we show pairwise correlations, we can see that the latter applies. In fact, all correlations  are significant but with low magnitude, suggesting that each of
the four signals of radicalization ($QC_{tweets}$, $QC_{profile}$, $C_{retweets}$, and $C_{lexical}$) characterizes a different dynamic of radicalization. As a consequence, we argue that it is not reasonable to combine these four dimensions into a single metric (e.g., by means of linear combinations, product, etc.).
Therefore, we keep these metrics separate and investigate whether these signals of radicalization can be used to identify distinct classes of behavior.

We leverage unsupervised learning techniques to cluster users based only on content-based and community-based metrics.
To analyze users that exhibit signals of engagement with QAnon, we retain only those with nonzero retweets and nonzero self-drafted tweets (947,597 out of 1.2M). 
Moreover, we assign a default value of zero to the metric $QC_{profile}$ to those with empty profile descriptions.

\begin{figure}[t]
\centering
\includegraphics[width=.65\linewidth]{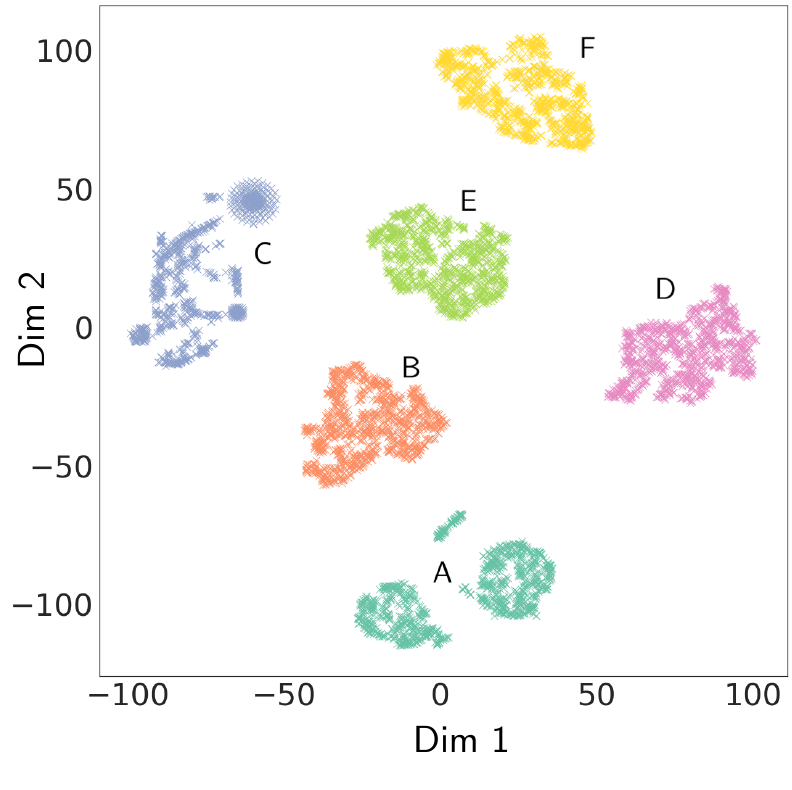}
\caption{Visualization of the 1,000 nearest neighbors to each cluster centroid using the t-SNE algorithm
}
\label{figure3r}
\end{figure}

We first employ the $k$-means clustering algorithm using Euclidean distance among the users in the 4-dimensional feature space composed of the four metrics.\footnote{As a robustness check, we repeated this procedure employing the cosine similarity and obtained almost identical results.} We leverage the library FAISS (Facebook AI Similarity Search) \cite{johnson2019billion} to run the $k$-means clustering algorithm. We select the number of clusters by varying $k$ in the range $[2,20]$ and identifying the elbow point at $k= 6$ clusters, which was further confirmed by the silhouette score. We then employ t-Distributed Stochastic Neighbor Embedding (t-SNE) \cite{van2008visualizing} to project users onto a two-dimensional space based on the four signals of radicalization. Figure \ref{figure3r} depicts the embedding results provided by t-SNE, where each point represents a user and colors indicate the clusters identified by the $k$-means algorithm. 
Six distinct clusters naturally emerge from the dimensionality reduction provided by t-SNE, which in turn corroborates the results obtained from the $k$-means algorithm.

\begin{figure*}[t]
\centering
\includegraphics[width=0.9\linewidth]{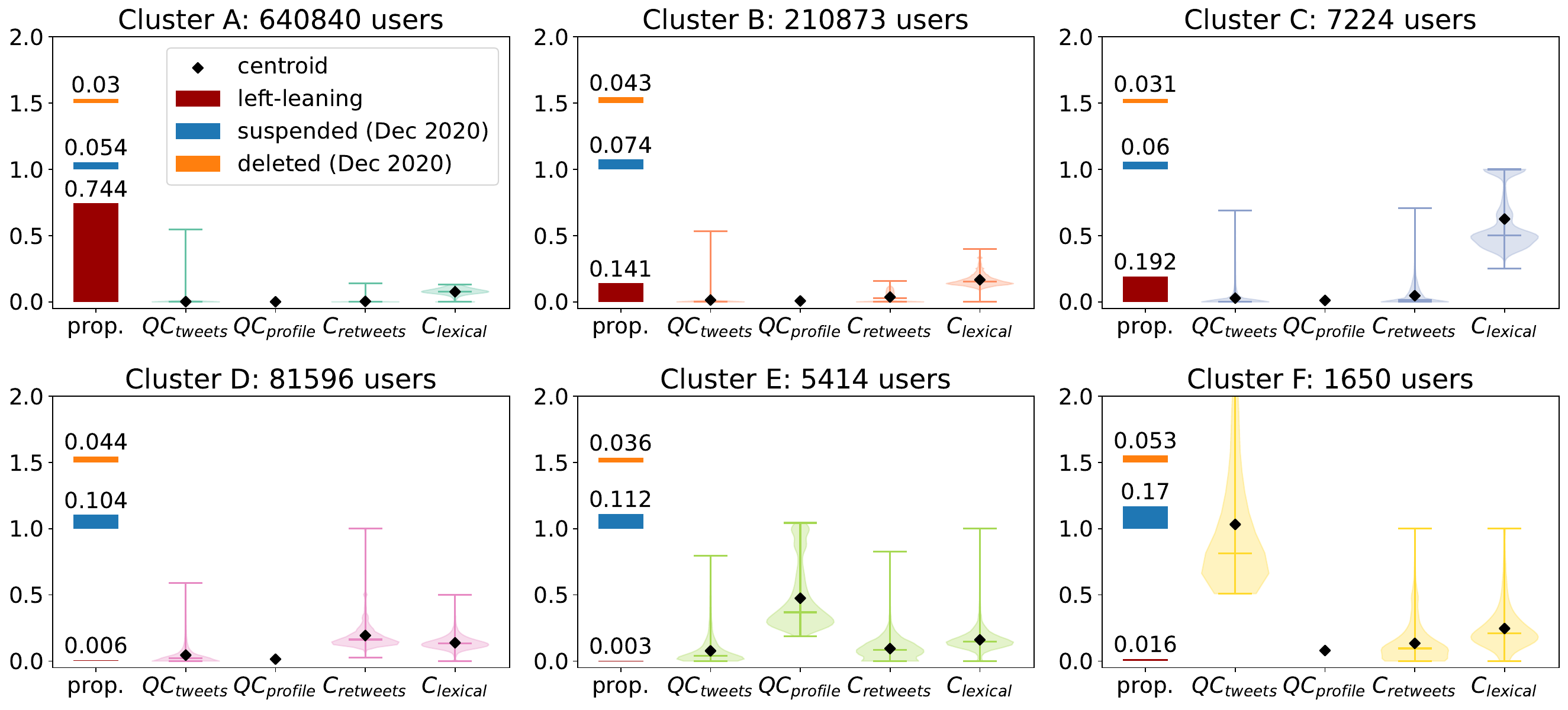}
\caption{
Distribution of the four metrics ($QC_{tweets}$, $QC_{profile}$, $C_{retweets}$, and $C_{lexical}$) for each cluster. The proportion of left-leaning, suspended, and deleted accounts within each cluster is also reported. Diamonds correspond to the values of the centroid in each cluster. Horizontal lines of the violin plot correspond to the minimum, median, and maximum values.
}
\label{figure1r}
\end{figure*}
 
\textbf{Finding \& Remarks}. Addressing \textit{RQ1}, we observe that six distinct clusters of behaviors naturally emerge when considering a set of four metrics to model radicalization processes. We conclude that the complex facets of radicalization cannot be modeled leveraging
a single metric.    

\subsection{Behavioral Classes of Radicalization (RQ2)}

\subsubsection{Characterizing Clusters}
To inspect whether these clusters can be tied to distinct behaviors, we 
examine the four signals of radicalization and users' sharing activity across the six clusters.

\begin{figure}[t]
\centering
\includegraphics[width=.55\linewidth]{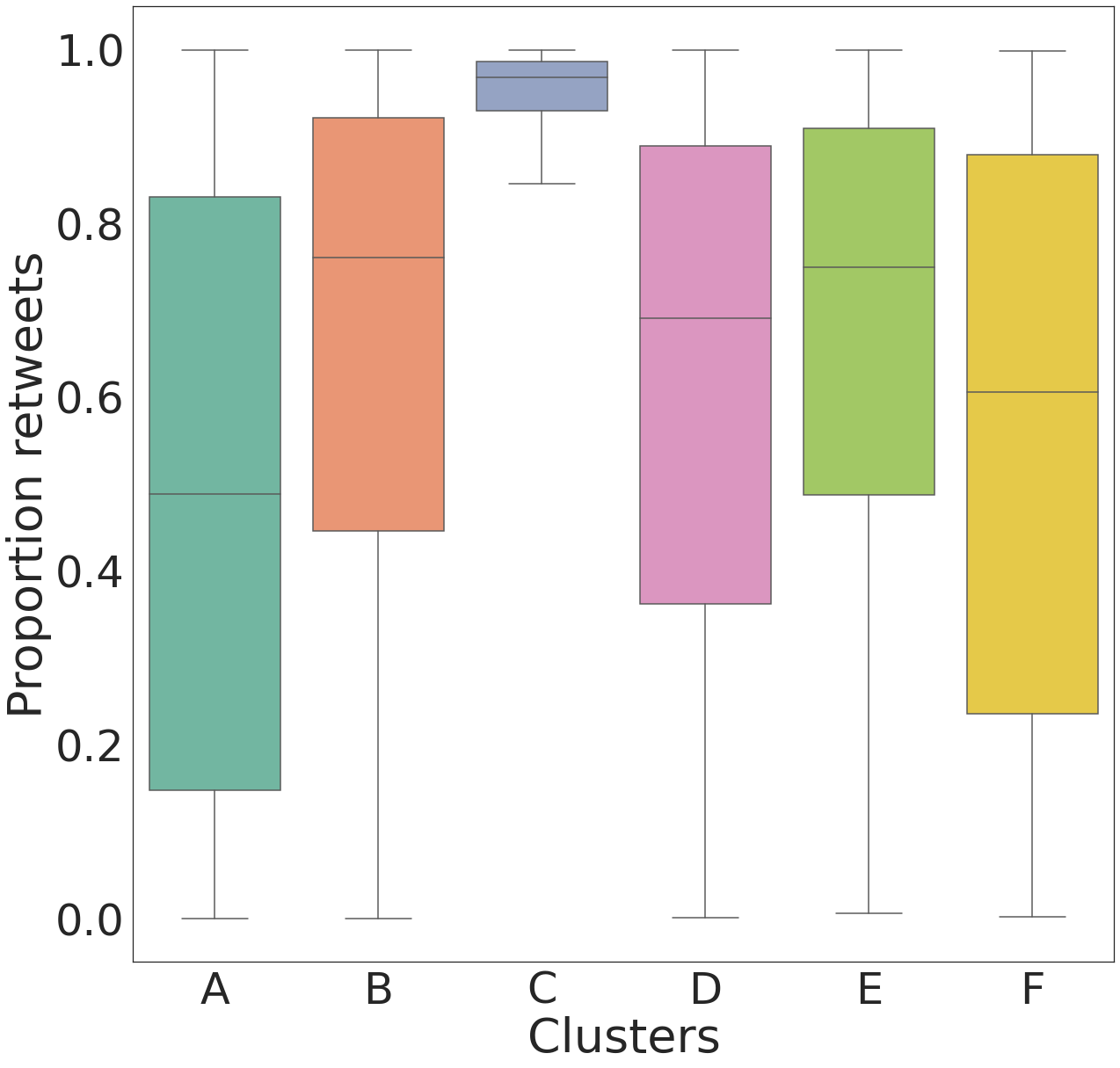}
\caption{
Distribution of the fraction of retweets over all tweets ($\frac{\textrm{No. of retweets}}{\textrm{No. of tweets}}$) of users in each cluster. Outliers are not shown. The upper whisker of the box plot is the largest data value within 1.5 IQR above the third quartile.
}
\label{figure2r} 
\end{figure}

\begin{figure}[t]
\centering
\includegraphics[width=.6\linewidth]{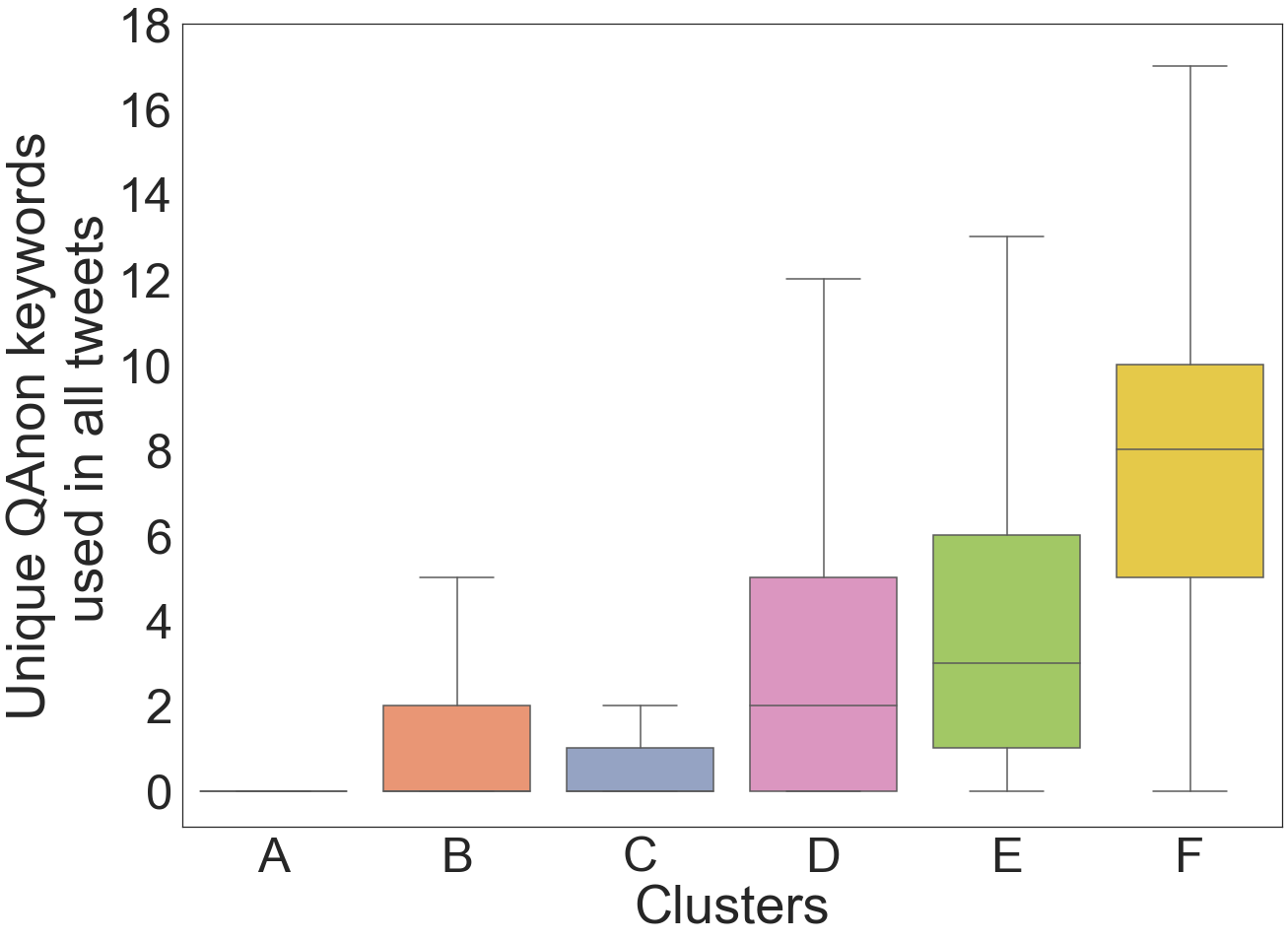}
\caption{
Unique number of QAnon keywords used by users in each cluster. Outliers not shown.
}
\label{figure5r}
\end{figure}

\begin{figure*}[t]
\centering
\includegraphics[width=\linewidth]{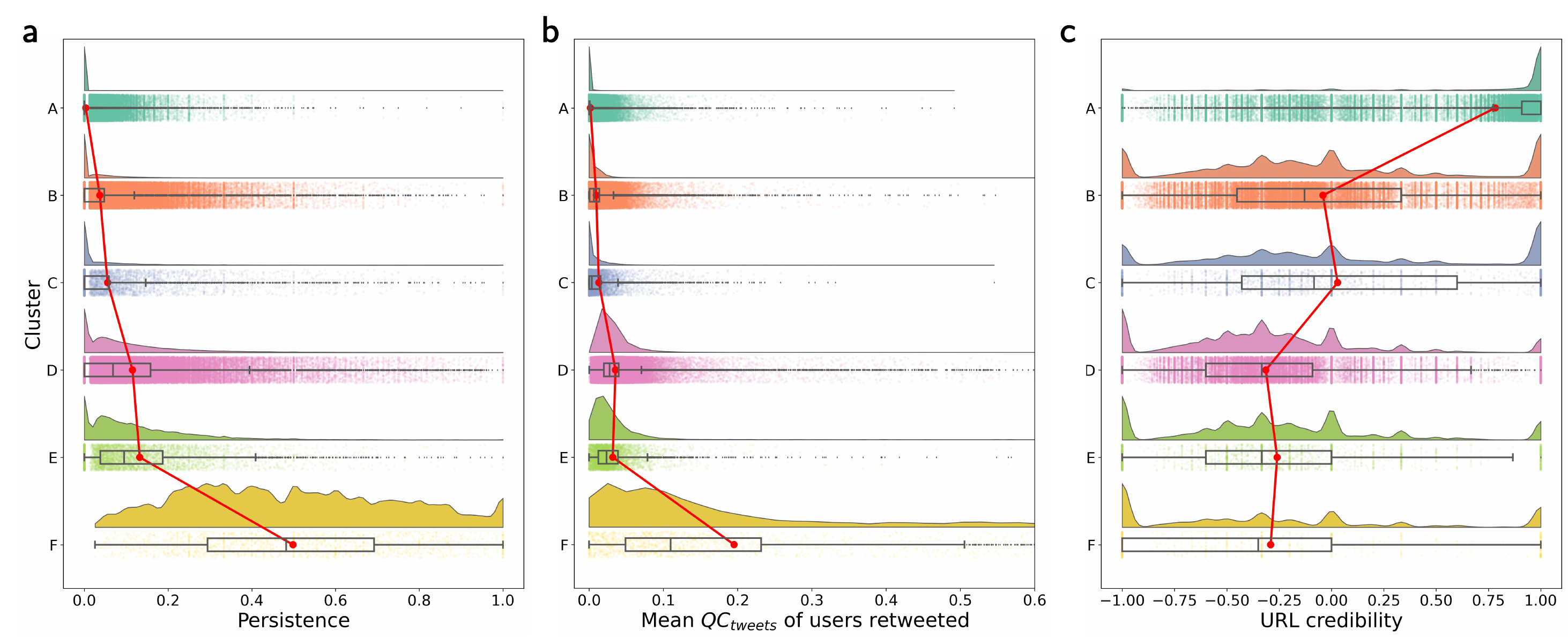}
\caption{
Clusters characteristics in terms of (a) Persistence,
(b) Mean $QC_{tweets}$ of retweeted users, 
(c) URL credibility 
}
\label{figure4r}
\end{figure*}

\begin{figure}[t]
\centering
\includegraphics[width=.6\linewidth]{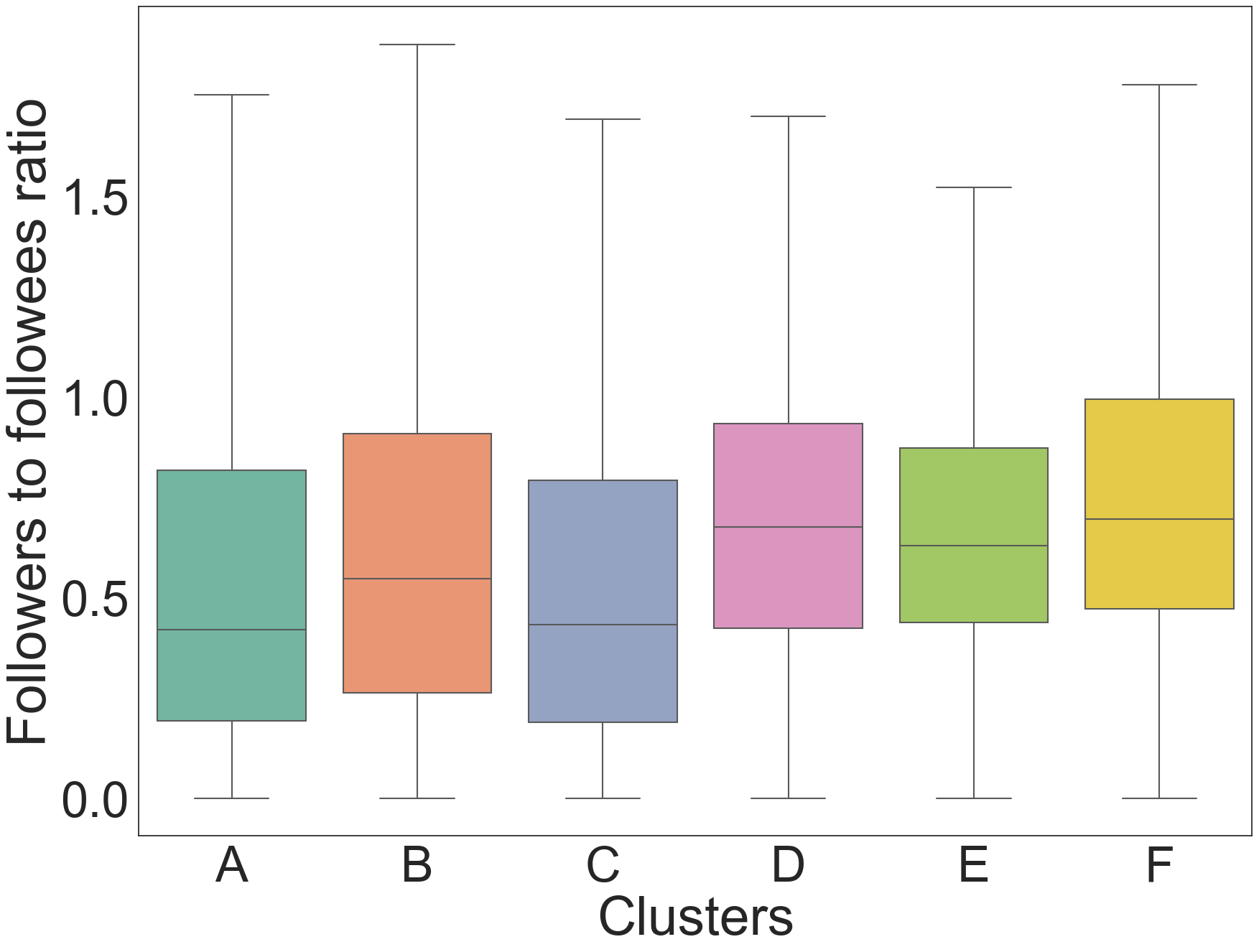}
\caption{
Distribution of $\frac{\textrm{No. of followers}}{\textrm{No. followees}}$ for users in each cluster. Outliers not shown.
}
\label{figure_follow} 
\end{figure}

\begin{figure}[t]
\centering
\includegraphics[width=\linewidth]{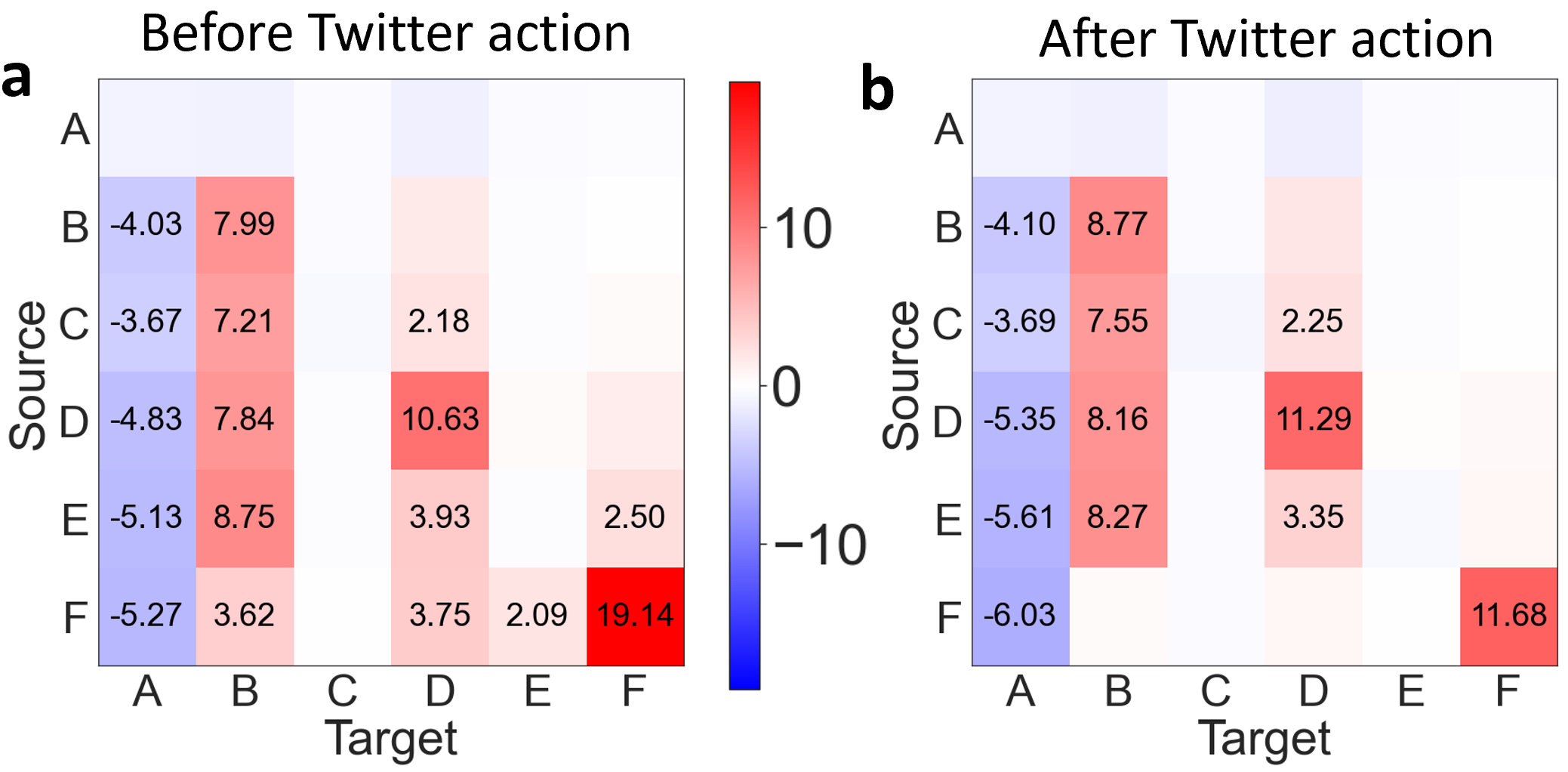}
\caption{
(a) Z-scores of observed retweets between clusters before Twitter action,
(b) Z-scores of observed retweets between clusters after Twitter action; values labeled are statistically significant (p-value $< 0.05$)
}
\label{figure8r}
\end{figure}

Figure \ref{figure1r} describes the six clusters by showing the distribution of their signals of radicalization as well as the proportion of left-leaning, suspended, and deleted accounts. The proportion of suspended and deleted accounts are calculated as of December 2020, which fully captures our period of analysis. 
Cluster A is characterized by low values of radicalization in every metric (median value of $QC_{tweets}$, $QC_{profile}$, and $C_{retweets}$ is 0), which can be expected given the high proportion of left-leaning users within the cluster. Similarly, Cluster B shows a low engagement, but it is comprised mostly of right-leaning users, including prominent political figures.
Despite showing a low level of engagement with QAnon content (low values for $QC_{tweets}$ and $C_{retweets}$), Cluster C exhibits very high lexical similarity with \emph{permanent QAnon users} ($C_{lexical}$). 

Delving deeper into the sharing activities of each cluster, in Figure \ref{figure2r} we depict the proportion of retweets over all tweets shared by the users within each cluster. We observe that users in Cluster C produce very few self-drafted tweets (high retweets over total tweets ratio). As $C_{lexical}$ is calculated only from users' self-drafted tweets, its high value for Cluster C might be due to the limited number of original content generated by users in this cluster.
Indeed, given the small sample of self-drafted tweets, $C_{lexical}$ might not properly capture the engagement with QAnon content. 

From Figure \ref{figure1r}, we notice that Cluster D displays low values in content-based metrics ($QC_{tweets}$ and $QC_{profile}$) but significant values for community-based metrics, especially $C_{retweets}$, which accounts for retweets to \emph{persistent QAnon users}.
For this reason, we refer to users in this cluster as \emph{amplifiers}. Cluster E stands out for high $QC_{profile}$ values, representing the cluster with the highest number of users declaring their support to QAnon in their profile description. Thus, we refer to them as \emph{self-declared} QAnon supporters. Finally, Cluster F exhibits low $QC_{profile}$, similar values to clusters D and E for community-based metrics ($C_{lexical}$ and $C_{retweets}$), and high values for $QC_{tweets}$, which indicates their massive activity in sharing QAnon content. 
We use the term \emph{hyper-active} QAnon promoters to refer to users in Cluster F. This term aligns with previous research \cite{luceri2021down} which found that hyper-active accounts are responsible for sharing the majority of tweets related to QAnon.

Overall, we find that the different clusters are tied with distinctive behaviors at varying levels of radicalization. 
For instance, those who explicitly specify QAnon keywords or URLs in their profile (Cluster E) are not the most active in promoting or amplifying QAnon content, 
whereas those who massively produce QAnon content (Cluster F) rarely state support for the conspiracy in their profile.  
Among these behavioral classes (note that we use \emph{class} and \emph{cluster} interchangeably), amplifiers (Cluster D), self-declared QAnon supporters (Cluster E), and hyper-active QAnon promoters (Cluster F) represent the most radicalized group of users, which is confirmed by the high proportions of suspended accounts and the low proportions of left-leaning users in these clusters. 

To further characterize these behavioral classes, we analyze in detail the engagement with QAnon content for each cluster by measuring (i) the total number of unique QAnon keywords employed over the period of observation and (ii) the users' \emph{persistence}, defined as the proportion of days in which the users shared QAnon content over the total number of days they were active on Twitter. More radicalized clusters tend to use a larger variety of keywords (Fig. \ref{figure5r}), indicating a broader involvement in QAnon-related topics (as opposed to specialist involvement with only a few keywords). Additionally, Figure \ref{figure4r}\textbf{a} shows the distribution of the users' \emph{persistence} for each cluster. Users in the most radicalized clusters (D, E, and F) tend to be more persistent in their QAnon involvement, indicating that more extreme users continuously supported QAnon despite Twitter intervention. By inspecting the users targeted in the retweet-sharing activity (i.e., who is the author of the re-shared tweet), we notice that hyper-active QAnon promoters (Cluster F) tend to endorse users who are highly engaged in QAnon content (Fig. \ref{figure4r}\textbf{b}).

Moreover, we look at the URLs shared by the users in their tweets and we compute the \emph{URL credibility} for each cluster as $\frac{(r - u)}{(r + u)}$, where $r$ is the number of reliable URLs shared by the users within the cluster and $u$ is the number of unreliable/conspiracy URLs shared by the users within the cluster.
We use the lists of 1,380 unreliable and conspiracy news sources (compiled from Media Bias/Fact Check, NewsGuard, and Zimdars) and 124 mainstream reliable news sources (compiled from Wikipedia) from \cite{sharma2022characterizing}. Figure \ref{figure4r}\textbf{c} depicts the distribution of the URL credibility for each cluster. As expected, the three most radicalized clusters (D, E, and F) share fewer reliable URLs with respect to the other clusters. 

Lastly, we investigate the possibility of a confounding variable within the identified clusters. We examine ``follow-back" patterns, where users agree to follow and constantly retweet each other. Pro-Trump follow-back communities have been observed \cite{torres2022manufacture}, and these users often have followers-to-followees ratios (FFRs) close to $1.0$. Although we find that the most radicalized clusters have higher median FFRs, it is important to note that the median ratios are all below $0.7$ (as shown in Fig. \ref{figure_follow}) and that the distributions are not significantly different. These findings suggest that although follow-back communities may have some impact on our analysis, such as QAnon supporters typically maintaining higher FFRs, none of our clusters are predominantly composed of follow-back users.

\subsubsection{Interactions Between Clusters}

We study the interactions between clusters by looking at the intra- and inter-cluster retweets exchanged by the users during the observation period. As cluster sizes are vastly different, we compare the observed number of retweets to a null model, which assumes that interactions occur by chance. Specifically, we randomize the cluster labels of the users, keeping cluster sizes constant, and compute the mean and standard deviation of the interactions between clusters for $1,000$ iterations. We then compute the z-scores to compare the observed retweets with the expected number of retweets from the null model. Figure \ref{figure8r} compares the z-scores of observed retweets before \textbf{(a)} and after \textbf{(b)} Twitter action against QAnon. In both periods, Cluster B received significantly more retweets than expected from other right-leaning clusters, while Cluster A received significantly fewer retweets than expected. This is reasonable since Cluster B includes several highly influential right-leaning politicians, while Cluster A contains mostly left-leaning users. Additionally, Clusters B, D, and F exhibit a significant number of intra-cluster retweets, which is a result that supports theoretical work on group polarization in radicalized communities \cite{yardi2010dynamic}.

After Twitter action, Cluster F limited their retweet activity and received fewer retweets from all clusters, especially from the most radicalized ones. As Cluster F contains users who explicitly use QAnon keywords and URLs in their tweets, this is consistent with previous findings \cite{sharma2022characterizing, cunningham2022network}, which highlighted how Twitter intervention discouraged users from expressing outright support to QAnon, and might also explain why users were hesitant to endorse hyper-active QAnon supporters (Cluster F). We suppose that users reduced their interactions with hyperactive QAnon supporters to minimize the risk of suspension rather than an inherent dislike of QAnon content or supporters. In fact, amplifiers (Cluster D) were not strongly affected by Twitter moderation, possibly due to lesser use of QAnon-related keywords (cf Fig. \ref{figure1r}), and continued receiving support from every class but the hyper-active QAnon supporters (Cluster F). 

\textbf{Findings \& Remarks}. In response to \textit{RQ2}, we discovered that the six distinct classes are associated with different behaviors, including archetypes such as hyper-active promoters and conspiracy amplifiers. Twitter mitigation strategies were effective to quell the former class but did not harm the latter, which kept receiving a significant volume of retweets.

\section{Discussion}


\subsection{Contributions}
In this paper, we studied the dynamics of Twitter users' radicalization with QAnon. We presented a methodological framework for measuring signals of radicalization that can be generalized to different platforms and fringe groups.
Leveraging a large-scale dataset of 240M tweets shared in the run-up to the 2020 US election, we show that radicalization is a multifaceted process that cannot be simply described by a unique feature. By looking at users' engagement with QAnon across different dimensions, we observed six distinct behavioral classes, some matching archetypes from theoretical studies, such as self-declared supporters and hyperactive promoters.
We studied the main characteristics of the six behavioral classes, finding that the most radicalized classes persistently share QAnon content and leverage low-credibility sources, 
spanning different topics and endorsing right-leaning political actors. Finally, we analyzed the interactions among behavioral classes in terms of retweets, observing a high volume of intra-class re-sharing messages, especially for amplifiers and hyperactive users. We discovered a reduced volume of retweets received by the latter class after Twitter moderation to demote QAnon narratives, while the former class was not significantly affected, possibly due to a more diverse sharing activity.

\subsection{Limitations}
There are a few limitations to our work. First, the detection of QAnon content based on keywords may be prone to errors and can result in false positives, as well as failure to capture messages that do not include keywords. 
Second, the lexical similarity metric $C_{lexical}$ might be affected by the sample size, especially for users with a few self-drafted tweets. 
However, the behavioral classes that originated from our clustering approach are clearly distinguished between users who only have high $C_{lexical}$ (i.e., Cluster C) versus users who have multiple high metrics (i.e., Cluster D), suggesting that the sample size does not significantly affect our conclusions. 
Third, we did not filter out automated accounts or detect inauthentic behaviors like political astroturf, a form of disinformation where participants in a coordinated campaign portray themselves as independent, ordinary citizens \cite{ratkiewicz2011truthy, ferrara2022twitter}, observed previously in QAnon-related activity on Twitter \cite{suresh2023tracking,dilley2022qanon}. Thus, we cannot discern whether accounts in our analysis correspond to real individuals or not, and some accounts observed in the analysis, particularly those in the most radicalized clusters, may have been part of a coordinated effort. 
Fourth, we observe a higher followers-to-followees ratio for the most radicalized clusters compared to the other clusters, which indicates that follow-back communities might influence some results, including rates of suspension, as the follow-back activity violates Twitter's usage norms.
Finally, we note that the methodology to infer political leanings, used to identify persistent users and characterize clusters, might not be completely accurate and thus affect our results.

\subsection{Conclusions and Future Work}

The results of this paper convey a two-pronged message. On the one hand, we observe that mitigation strategies adopted by social media platforms can be effective to quell hyperactive users spreading conspiracies. On the other hand, we show that less evident radicalized behaviors can escape moderation, leading to the persistence of problematic content in online platforms.
Based on these premises, our findings can inform social media providers, regulators, and policymakers to formulate strategies to counter the circulation of conspiracy theories and fringe narratives on social media.

Our findings show that radicalization within a given conspiracy on social media platforms can generate diverse behaviors, which should be observed through the lens of a heterogeneous set of indicators. The signals of radicalization presented in this work are a primary example of the diverse facets that can be captured to model radicalization processes. Indeed, our methodological framework is not comprehensive and  can be augmented by encompassing a larger variety of metrics. This work represents the first building block in the large-scale modeling of radicalization within fringe communities and can pave the way to research validating and augmenting our methodology. In the future, we aim to develop new metrics quantifying beyond the ``Network" stage proposed in radicalization theories, i.e., ``Need'' and ``Narrative''. 
These metrics may allow us to include factors such as anxiety, isolation, and anger, thus enhancing the identification of topics and opinions in users' messages. 

And there might be other avenues for future work. For example, one could apply our framework to different extreme groups and other platforms, which could help to further our understanding of conspiracy and fringe groups in other settings. This would also lead to assessing differences and similarities between fringe communities acting on different platforms. Moreover, our methodology could be leveraged in longitudinal studies that aim to investigate radicalization patterns over time, even in relation to real-world events. 
For example, we aim to extend this study to periods of social upheaval or unrest, such as the January 6, 2021 Capitol Hill riots, to investigate the possible link between conspiracy rhetoric on social media and real-world events.

We end by noting that content moderation and user bans appear to reduce conspiracy content, but they may be counterproductive, leading users to migrate to platforms with lower diversity of ideas and no moderation against problematic content \cite{ali2021understanding, pierri2022does, cinelli2022conspiracy}. 
To moderate divisive issues, actions like diversifying content and user recommendations may be more far-reaching, effective, and fair.


\subsection{Broader perspective, ethics and competing interests}
In this study, we do not identify individual users or provide exact quotes from users and present data in an aggregated manner in order to protect user privacy. For the same reason, we do not share the inferred political leanings or the list of persistent QAnon users. The Tweet IDs of data used in this paper can however be accessed via the original public dataset \cite{chen2022election2020} in accordance with Twitter's data-sharing policy. We caution applying our findings to censor users---content moderation against accounts should remain impartial and consistent with Twitter's rules.
Project approved by USC IRB (\#UP-21-00005-AM001).

\section*{Acknowledgements}
Work supported in part by
NSF (grant \#2051101) and PRIN grant HOPE (FP6, Italian Ministry of Education).
This work would not have been possible without the insights and guidance from Karishma Sharma and support from Emily Chen.

\bibliography{aaai23}

\begin{thebibliography}{53}
\providecommand{\natexlab}[1]{#1}

\bibitem[{Abilov et~al.(2021)Abilov, Hua, Matatov, Amir, and
  Naaman}]{abilov2021voterfraud2020}
Abilov, A.; Hua, Y.; Matatov, H.; Amir, O.; and Naaman, M. 2021.
\newblock VoterFraud2020: a Multi-modal Dataset of Election Fraud Claims on
  Twitter.
\newblock In \emph{Proceedings of the International AAAI Conference on Web and
  Social Media}, volume~15, 901--912.

\bibitem[{Ali et~al.(2021)Ali, Saeed, Aldreabi, Blackburn, De~Cristofaro,
  Zannettou, and Stringhini}]{ali2021understanding}
Ali, S.; Saeed, M.~H.; Aldreabi, E.; Blackburn, J.; De~Cristofaro, E.;
  Zannettou, S.; and Stringhini, G. 2021.
\newblock Understanding the effect of deplatforming on social networks.
\newblock In \emph{13th ACM Web Science Conference 2021}, 187--195.

\bibitem[{Amarasingam and Argentino(2020)}]{amarasingam2020qanon}
Amarasingam, A.; and Argentino, M.-A. 2020.
\newblock The QAnon conspiracy theory: A security threat in the making.
\newblock \emph{CTC Sentinel}, 13(7): 37--44.

\bibitem[{B{\'e}langer et~al.(2019)B{\'e}langer, Moyano, Muhammad, Richardson,
  Lafreni{\`e}re, McCaffery, Framand, and Nociti}]{belanger2019radicalization}
B{\'e}langer, J.~J.; Moyano, M.; Muhammad, H.; Richardson, L.; Lafreni{\`e}re,
  M.-A.~K.; McCaffery, P.; Framand, K.; and Nociti, N. 2019.
\newblock Radicalization leading to violence: A test of the 3N model.
\newblock \emph{Frontiers in psychiatry}, 10: 42.

\bibitem[{Bird, Klein, and Loper(2009)}]{bird2009natural}
Bird, S.; Klein, E.; and Loper, E. 2009.
\newblock \emph{Natural language processing with Python: analyzing text with
  the natural language toolkit}.
\newblock " O'Reilly Media, Inc.".

\bibitem[{Chen, Deb, and Ferrara(2022)}]{chen2022election2020}
Chen, E.; Deb, A.; and Ferrara, E. 2022.
\newblock \# Election2020: The first public Twitter dataset on the 2020 US
  Presidential election.
\newblock \emph{Journal of Computational Social Science}, 5(1).

\bibitem[{Cinelli et~al.(2022)Cinelli, Etta, Avalle, Quattrociocchi, Di~Marco,
  Valensise, Galeazzi, and Quattrociocchi}]{cinelli2022conspiracy}
Cinelli, M.; Etta, G.; Avalle, M.; Quattrociocchi, A.; Di~Marco, N.; Valensise,
  C.; Galeazzi, A.; and Quattrociocchi, W. 2022.
\newblock Conspiracy theories and social media platforms.
\newblock \emph{Current Opinion in Psychology}, 101407.

\bibitem[{Cunningham and Everton(2022)}]{cunningham2022network}
Cunningham, D.; and Everton, S. 2022.
\newblock A Network Analysis of Twitter's Crackdown on the QAnon Conversation.
\newblock \emph{Journal of Social Structure}, 23(1): 4--27.

\bibitem[{De~Zeeuw et~al.(2020)De~Zeeuw, Hagen, Peeters, and
  Jokubauskaite}]{de2020tracing}
De~Zeeuw, D.; Hagen, S.; Peeters, S.; and Jokubauskaite, E. 2020.
\newblock Tracing normiefication: A cross-platform analysis of the QAnon
  conspiracy theory.
\newblock \emph{First Monday}.

\bibitem[{Dilley, Welna, and Foster(2022)}]{dilley2022qanon}
Dilley, L.; Welna, W.; and Foster, F. 2022.
\newblock QAnon Propaganda on Twitter as Information Warfare: Influencers,
  Networks, and Narratives.
\newblock arXiv:2207.05118.

\bibitem[{Fabbri et~al.(2022)Fabbri, Wang, Bonchi, Castillo, and
  Mathioudakis}]{fabbri2022rewiring}
Fabbri, F.; Wang, Y.; Bonchi, F.; Castillo, C.; and Mathioudakis, M. 2022.
\newblock Rewiring What-to-Watch-Next Recommendations to Reduce Radicalization
  Pathways.
\newblock In \emph{Proceedings of the ACM Web Conference 2022}, 2719--2728.

\bibitem[{Fernandez, Asif, and Alani(2018)}]{fernandez2018understanding}
Fernandez, M.; Asif, M.; and Alani, H. 2018.
\newblock Understanding the roots of radicalisation on twitter.
\newblock In \emph{Proceedings of the 10th ACM conference on web science},
  1--10.

\bibitem[{Ferrara(2020)}]{ferrara2020what}
Ferrara, E. 2020.
\newblock What types of COVID-19 conspiracies are populated by Twitter bots?
\newblock \emph{First Monday}, 25(6).

\bibitem[{Ferrara(2022)}]{ferrara2022twitter}
Ferrara, E. 2022.
\newblock Twitter spam and false accounts prevalence, detection, and
  characterization: A survey.
\newblock \emph{First Monday}.

\bibitem[{Hanley, Kumar, and Durumeric(2022)}]{hanley2022no}
Hanley, H.~W.; Kumar, D.; and Durumeric, Z. 2022.
\newblock No Calm in The Storm: Investigating QAnon Website Relationships.
\newblock In \emph{Proceedings of the International AAAI Conference on Web and
  Social Media}, volume~16, 299--310.

\bibitem[{Hannah(2021)}]{hannah2021qanon}
Hannah, M. 2021.
\newblock QAnon and the information dark age.
\newblock \emph{First Monday}.

\bibitem[{Hine et~al.(2017)Hine, Onaolapo, De~Cristofaro, Kourtellis,
  Leontiadis, Samaras, Stringhini, and Blackburn}]{hine2017kek}
Hine, G.~E.; Onaolapo, J.; De~Cristofaro, E.; Kourtellis, N.; Leontiadis, I.;
  Samaras, R.; Stringhini, G.; and Blackburn, J. 2017.
\newblock Kek, cucks, and god emperor trump: A measurement study of 4chan’s
  politically incorrect forum and its effects on the web.
\newblock In \emph{11th International AAAI Conference on Web and Social Media}.

\bibitem[{Imhoff, Dieterle, and Lamberty(2021)}]{imhoff2021resolving}
Imhoff, R.; Dieterle, L.; and Lamberty, P. 2021.
\newblock Resolving the puzzle of conspiracy worldview and political activism:
  Belief in secret plots decreases normative but increases nonnormative
  political engagement.
\newblock \emph{Social Psychological and Personality Science}, 12(1): 71--79.

\bibitem[{Johnson, Douze, and J{\'e}gou(2019)}]{johnson2019billion}
Johnson, J.; Douze, M.; and J{\'e}gou, H. 2019.
\newblock Billion-scale similarity search with {GPUs}.
\newblock \emph{IEEE Transactions on Big Data}, 7(3): 535--547.

\bibitem[{Jolley and Paterson(2020)}]{jolley2020pylons}
Jolley, D.; and Paterson, J.~L. 2020.
\newblock Pylons ablaze: Examining the role of 5G COVID-19 conspiracy beliefs
  and support for violence.
\newblock \emph{British journal of social psychology}, 59(3).

\bibitem[{Kruglanski et~al.(2014)Kruglanski, Gelfand, B{\'e}langer, Sheveland,
  Hetiarachchi, and Gunaratna}]{kruglanski2014psychology}
Kruglanski, A.~W.; Gelfand, M.~J.; B{\'e}langer, J.~J.; Sheveland, A.;
  Hetiarachchi, M.; and Gunaratna, R. 2014.
\newblock The psychology of radicalization and deradicalization: How
  significance quest impacts violent extremism.
\newblock \emph{Political Psychology}, 35: 69--93.

\bibitem[{Kruglanski et~al.(2022)Kruglanski, Molinario, Ellenberg, and
  Di~Cicco}]{kruglanski2022terrorism}
Kruglanski, A.~W.; Molinario, E.; Ellenberg, M.; and Di~Cicco, G. 2022.
\newblock Terrorism and conspiracy theories: A view from the 3N model of
  radicalization.
\newblock \emph{Current Opinion in Psychology}, 101396.

\bibitem[{Luceri, Cardoso, and Giordano(2021)}]{luceri2021down}
Luceri, L.; Cardoso, F.; and Giordano, S. 2021.
\newblock Down the bot hole: Actionable insights from a one-year analysis of
  bot activity on Twitter.
\newblock \emph{First Monday}.

\bibitem[{Luceri, Cresci, and Giordano(2021)}]{luceri2021social}
Luceri, L.; Cresci, S.; and Giordano, S. 2021.
\newblock Social Media against Society.
\newblock \emph{The Internet and the 2020 Campaign}, 1.

\bibitem[{McPherson, Smith-Lovin, and Cook(2001)}]{mcpherson2001birds}
McPherson, M.; Smith-Lovin, L.; and Cook, J.~M. 2001.
\newblock Birds of a feather: Homophily in social networks.
\newblock \emph{Annual review of sociology}, 415--444.

\bibitem[{Metaxas et~al.(2015)Metaxas, Mustafaraj, Wong, Zeng, O'Keefe, and
  Finn}]{metaxas2015retweets}
Metaxas, P.; Mustafaraj, E.; Wong, K.; Zeng, L.; O'Keefe, M.; and Finn, S.
  2015.
\newblock What do retweets indicate? Results from user survey and meta-review
  of research.
\newblock In \emph{Proceedings of the international AAAI conference on web and
  social media}, 658--661.

\bibitem[{Monroe, Colaresi, and Quinn(2008)}]{monroe2008fightin}
Monroe, B.~L.; Colaresi, M.~P.; and Quinn, K.~M. 2008.
\newblock Fightin'words: Lexical feature selection and evaluation for
  identifying the content of political conflict.
\newblock \emph{Political Analysis}, 16(4): 372--403.

\bibitem[{Neo(2019)}]{neo2019internet}
Neo, L.~S. 2019.
\newblock An Internet-mediated pathway for online radicalisation: RECRO.
\newblock In \emph{Violent Extremism: Breakthroughs in Research and Practice},
  62--89. IGI Global.

\bibitem[{Nogara et~al.(2022)Nogara, Vishnuprasad, Cardoso, Ayoub, Giordano,
  and Luceri}]{nogara2022disinformation}
Nogara, G.; Vishnuprasad, P.~S.; Cardoso, F.; Ayoub, O.; Giordano, S.; and
  Luceri, L. 2022.
\newblock The disinformation dozen: An exploratory analysis of covid-19
  disinformation proliferation on twitter.
\newblock In \emph{14th ACM Web Science Conference 2022}, 348--358.

\bibitem[{Nouh, Nurse, and Goldsmith(2019)}]{nouh2019understanding}
Nouh, M.; Nurse, J.~R.; and Goldsmith, M. 2019.
\newblock Understanding the radical mind: Identifying signals to detect
  extremist content on twitter.
\newblock In \emph{2019 IEEE International Conference on Intelligence and
  Security Informatics}, 98--103.

\bibitem[{Papakyriakopoulos, Serrano, and
  Hegelich(2020)}]{papakyriakopoulos2020spread}
Papakyriakopoulos, O.; Serrano, J. C.~M.; and Hegelich, S. 2020.
\newblock The spread of COVID-19 conspiracy theories on social media and the
  effect of content moderation.
\newblock \emph{The Harvard Kennedy School (HKS) Misinformation Review}, 18.

\bibitem[{Papasavva et~al.(2021)Papasavva, Blackburn, Stringhini, Zannettou,
  and Cristofaro}]{papasavva2021qoincidence}
Papasavva, A.; Blackburn, J.; Stringhini, G.; Zannettou, S.; and Cristofaro,
  E.~D. 2021.
\newblock “Is it a qoincidence?”: An exploratory study of QAnon on Voat.
\newblock In \emph{Proceedings of the Web Conference 2021}, 460--471.

\bibitem[{Phadke, Samory, and Mitra(2022)}]{phadke2022pathways}
Phadke, S.; Samory, M.; and Mitra, T. 2022.
\newblock Pathways through Conspiracy: The Evolution of Conspiracy
  Radicalization through Engagement in Online Conspiracy Discussions.
\newblock In \emph{Proceedings of the International AAAI Conference on Web and
  Social Media}, 770--781.

\bibitem[{Pierri et~al.(2023)Pierri, DeVerna, Yang, Axelrod, Bryden, and
  Menczer}]{pierri2022one}
Pierri, F.; DeVerna, M.~R.; Yang, K.-C.; Axelrod, D.; Bryden, J.; and Menczer,
  F. 2023.
\newblock One Year of COVID-19 Vaccine Misinformation on Twitter: Longitudinal
  Study.
\newblock \emph{Journal of Medical Internet Research}, 25: e42227.

\bibitem[{Pierri, Luceri, and Ferrara(2022)}]{pierri2022does}
Pierri, F.; Luceri, L.; and Ferrara, E. 2022.
\newblock How does Twitter account moderation work? Dynamics of account
  creation and suspension during major geopolitical events.
\newblock \emph{arXiv preprint arXiv:2209.07614}.

\bibitem[{Ratkiewicz et~al.(2011)Ratkiewicz, Conover, Meiss, Gon{\c{c}}alves,
  Patil, Flammini, and Menczer}]{ratkiewicz2011truthy}
Ratkiewicz, J.; Conover, M.; Meiss, M.; Gon{\c{c}}alves, B.; Patil, S.;
  Flammini, A.; and Menczer, F. 2011.
\newblock Truthy: mapping the spread of astroturf in microblog streams.
\newblock In \emph{Proceedings of the 20th international conference companion
  on World wide web}, 249--252.

\bibitem[{Ribeiro et~al.(2020)Ribeiro, Ottoni, West, Almeida, and
  Meira~Jr}]{ribeiro2020auditing}
Ribeiro, M.~H.; Ottoni, R.; West, R.; Almeida, V.~A.; and Meira~Jr, W. 2020.
\newblock Auditing radicalization pathways on YouTube.
\newblock In \emph{Proceedings of the 2020 conference on fairness,
  accountability, and transparency}, 131--141.

\bibitem[{Rowe and Saif(2016)}]{rowe2016mining}
Rowe, M.; and Saif, H. 2016.
\newblock Mining pro-ISIS radicalisation signals from social media users.
\newblock In \emph{tenth international AAAI conference on web and social
  media}.

\bibitem[{Sageman(2008)}]{sageman2008strategy}
Sageman, M. 2008.
\newblock A strategy for fighting international Islamist terrorists.
\newblock \emph{The Annals of the American Academy of Political and Social
  Science}, 618(1): 223--231.

\bibitem[{Schulze et~al.(2022)Schulze, Hohner, Greipl, Girgnhuber, Desta, and
  Rieger}]{schulze2022far}
Schulze, H.; Hohner, J.; Greipl, S.; Girgnhuber, M.; Desta, I.; and Rieger, D.
  2022.
\newblock Far-right conspiracy groups on fringe platforms: a longitudinal
  analysis of radicalization dynamics on Telegram.
\newblock \emph{Convergence}, 13548565221104977.

\bibitem[{Sharma, Ferrara, and Liu(2022)}]{sharma2022characterizing}
Sharma, K.; Ferrara, E.; and Liu, Y. 2022.
\newblock Characterizing Online Engagement with Disinformation and Conspiracies
  in the 2020 US Presidential Election.
\newblock In \emph{Proceedings of the International AAAI Conference on Web and
  Social Media}, 908--919.

\bibitem[{Stecula and Pickup(2021)}]{stecula2021social}
Stecula, D.~A.; and Pickup, M. 2021.
\newblock Social media, cognitive reflection, and conspiracy beliefs.
\newblock \emph{Frontiers in Political Science}, 3: 647957.

\bibitem[{Stella, Ferrara, and De~Domenico(2018)}]{stella2018bots}
Stella, M.; Ferrara, E.; and De~Domenico, M. 2018.
\newblock Bots increase exposure to negative and inflammatory content in online
  social systems.
\newblock \emph{Proceedings of the National Academy of Sciences}, 115(49):
  12435--12440.

\bibitem[{Suresh et~al.(2023)Suresh, Nogara, Cardoso, Cresci, Giordano, and
  Luceri}]{suresh2023tracking}
Suresh, V.~P.; Nogara, G.; Cardoso, F.; Cresci, S.; Giordano, S.; and Luceri,
  L. 2023.
\newblock Tracking Fringe and Coordinated Activity on Twitter Leading Up To the
  US Capitol Attack.
\newblock \emph{arXiv preprint arXiv:2302.04450}.

\bibitem[{Tollefson(2021)}]{tollefson2021tracking}
Tollefson, J. 2021.
\newblock Tracking QAnon: how Trump turned conspiracy-theory research upside
  down.
\newblock \emph{Nature}, 590(7845).

\bibitem[{Torok(2013)}]{torok2013developing}
Torok, R. 2013.
\newblock Developing an explanatory model for the process of online
  radicalisation and terrorism.
\newblock \emph{Security Informatics}, 2(1): 1--10.

\bibitem[{Torres-Lugo, Yang, and Menczer(2022)}]{torres2022manufacture}
Torres-Lugo, C.; Yang, K.-C.; and Menczer, F. 2022.
\newblock The Manufacture of Partisan Echo Chambers by Follow Train Abuse on
  Twitter.
\newblock In \emph{Proceedings of the International AAAI Conference on Web and
  Social Media}, volume~16, 1017--1028.

\bibitem[{Van~der Maaten and Hinton(2008)}]{van2008visualizing}
Van~der Maaten, L.; and Hinton, G. 2008.
\newblock Visualizing data using t-SNE.
\newblock \emph{Journal of machine learning research}, 9(11).

\bibitem[{Vegetti and Littvay(2022)}]{vegetti2022belief}
Vegetti, F.; and Littvay, L. 2022.
\newblock Belief in conspiracy theories and attitudes toward political
  violence.
\newblock \emph{Italian Political Science Review/Rivista Italiana di Scienza
  Politica}, 52(1).

\bibitem[{Xu, Sasahara et~al.(2022)}]{xu2022network}
Xu, W.; Sasahara, K.; et~al. 2022.
\newblock A network-based approach to QAnon user dynamics and topic diversity
  during the COVID-19 infodemic.
\newblock \emph{APSIPA Transactions on Signal and Information Processing},
  11(1).

\bibitem[{Yardi and Boyd(2010)}]{yardi2010dynamic}
Yardi, S.; and Boyd, D. 2010.
\newblock Dynamic debates: An analysis of group polarization over time on
  twitter.
\newblock \emph{Bulletin of science, technology \& society}, 30(5): 316--327.

\bibitem[{Zannettou et~al.(2018)Zannettou, Caulfield, Blackburn, De~Cristofaro,
  Sirivianos, Stringhini, and Suarez-Tangil}]{zannettou2018origins}
Zannettou, S.; Caulfield, T.; Blackburn, J.; De~Cristofaro, E.; Sirivianos, M.;
  Stringhini, G.; and Suarez-Tangil, G. 2018.
\newblock On the origins of memes by means of fringe web communities.
\newblock In \emph{Proceedings of the Internet Measurement Conference 2018},
  188--202.

\bibitem[{Zihiri et~al.(2022)Zihiri, Lima, Han, Cha, and Lee}]{zihiri2022qanon}
Zihiri, S.; Lima, G.; Han, J.; Cha, M.; and Lee, W. 2022.
\newblock QAnon shifts into the mainstream, remains a far-right ally.
\newblock \emph{Heliyon}, 8(2): e08764.

\end{thebibliography}

\end{document}